# Surface Plasmonic Lattice Solitons in Semi-infinite Graphene Sheet Arrays

Zhouqing Wang, Bing Wang, Hua Long, Kai Wang, and Peixiang Lu

*Abstract*—We investigate the surface plasmonic lattice solitons (PLSs) in semi-infinite graphene sheet arrays. The surface soliton is formed as the SPPs tunneling is inhibited by the graphene nonlinearity, and meanwhile the incident power should be above a threshold value. Thanks to the strong confinement of surface plasmon polaritons (SPPs) on graphene, the effective width of surface PLSs can be squeezed into deep-subwavelength scale of ~ $0.001\lambda$. Based on the stable propagation of surface PLSs, we find that the light propagation can be switched from the array boundary to the inner graphene sheets by reducing the incident power or increasing the chemical potential of graphene. The study may find promising application in optical switches on deep-subwavelength scale.

## I. Introduction

The optical phenomena supported by the periodic discrete structures such as dielectric waveguide arrays and plasmonic lattices have attracted great interest [1-3]. These periodic discrete structures provide a fertile platform for the light to exhibit various novel properties. As the optical phenomena take place with nonlinear effects considered, the light could exhibit more interesting behaviours, including diffraction-free propagation and all-optical control [3-5]. The discrete soliton is a typical class of the nonlinear optical phenomena in the periodic discrete structure [6-9]. To realize extremely small width of solitons with relatively low input power, ones have paid attention to the solitons formed in plasmonic lattice such as the metal-dielectric arrays instead of dielectric lattices [10-13]. The surface plasmon polaritons (SPPs) on metals manifest strong confinement of the light field, which could make the nonlinear effect stimulated readily. In plasmonic lattice, the discrete solitons are known as plasmonic lattice solitons (PLSs). The solitons are formed as the SPPs tunneling is inhibited by nonlinear effect. Usually PLSs are investigated in the infinite (or homogenous) plasmonic lattice [12-14]. There is also a type of PLSs existing at the boundary of the truncated plasmonic lattice, which is the surface plasmonic lattice soliton [15]. Compared with the counterparts in the infinite plasmonic lattice, the surface PLSs are unique which require the input power above a threshold value.

Many analogous nonlinear optical phenomena including the discrete solitons, have also been studied in the graphene metamaterials [16-20]. In comparison with metals, the SPPs on graphene have superior features including the stronger confinement, more flexible tunability, and lower propagation loss [21-24]. Apart from that, graphene itself exhibits strong optical nonlinearity [20,25,26]. These features make graphene sheet arrays an ideal substitution of the metallic waveguide arrays for researching plasmonic lattice solitons. In this letter, we systematically investigate the surface PLSs in the semi-infinite graphene sheet arrays (GSAs). The structure is composed of semi-infinite graphene sheets periodically embedding in the host linear dielectric. At the boundary of the semi-infinite graphene sheet arrays, the graphene nonlinearity excited by the light power above a threshold value balances the SPPs tunneling to form the surface PLSs. Due to the strong confinement of SPPs on graphene, the transverse distribution of the surface PLSs can be compressed into deep-subwavelength scale (~ $0.001\lambda$). The threshold power of the surface PLS in the semi-infinite GSAs is also analyzed. More interestingly, as we enhance the chemical potential or reduce the input power, the stable propagation of surface PLSs will be interrupted and the light beam will diffract from the boundary to the inner graphene sheets. At a fixed distance, the output position of light can be manipulated flexibly by tuning the chemical potential and the input power. The study may find potential applications in the optical switches and optical circuits on deep-subwavelength scale.

## II. Mode Distribution and Propagation of Surface PLSs

We firstly investigate the transverse field distribution and propagation of surface PLSs in the semi-infinite graphene sheet arrays. The structure is shown in Fig. 1, where the semi-infinite graphene sheet arrays ($x > 0$) embedding in the linear dielectric ($\varepsilon_d = 2.25$). The graphene sheets are represented by the black lines and the distance between adjacent graphene sheets is $d = 40$ nm.

Considering the transverse magnetic polarization (TM), we transform the Maxwell's equations into the matrix form [12]

$$\begin{pmatrix} 0 & \dfrac{k_0 \varepsilon_r(x)}{\eta_0} \\ \dfrac{\eta_0}{k_0} \dfrac{\partial}{\partial x} \dfrac{1}{\varepsilon_r(x)} \dfrac{\partial}{\partial x} + k_0 \eta_0 & 0 \end{pmatrix} \begin{pmatrix} H_y \\ E_x \end{pmatrix} = k_z \begin{pmatrix} H_y \\ E_x \end{pmatrix}, \quad (1)$$

where $\varepsilon_r(x)$ stands for the relative permittivity along the $x$ axis. Equation (1) is aimed at solving the eigen problem. The transverse magnetic field and electric field, $H_y$ and $E_x$, compose of the eigenvector, while the propagation constant $k_z$ is the eigenvalue. In the calculation, graphene is treated as a thin film with an equivalent thickness $\Delta \approx 1$ nm. Then the relative equivalent permittivity of graphene could be given by $\varepsilon_g = 1 + i\sigma_g \eta_0/(k_0 \Delta)$ [16], where $\eta_0$ and $k_0$ are the impendence and propagation constant in vacuum. The nonlinearity of graphene originates from its surface conductivity which can be written as $\sigma_g = \sigma_{g,linear} + \sigma^{NL}|E_z|^2$. The nonlinear conductivity $\sigma^{NL}$ is given by [17,25]

$$\sigma^{NL} = -i\frac{3}{8}\frac{e^2}{\pi\hbar^2}(\frac{eV_F}{\mu_c\omega})^2\frac{\mu_c}{\omega}, \quad (2)$$

where $\omega$ denotes the frequency and the Fermi velocity $V_F \approx c/300$. It is obvious that the equivalent permittivity of graphene is intensity-dependent. The linear part of the surface conductivity $\sigma_{g,linear}$ ($\lambda$, $\mu_c$, $\tau$) is governed by the Kubo formula [27,28], where $\lambda$ is the incident wavelength in air, $\mu_c$ is the chemical potential of graphene, and $\tau$ is the momentum relaxation time. Here, these parameters are respectively set as $\lambda = 10$ μm, $\mu_c = 0.15$ eV, $\tau = 0.5$ ps [29,30].

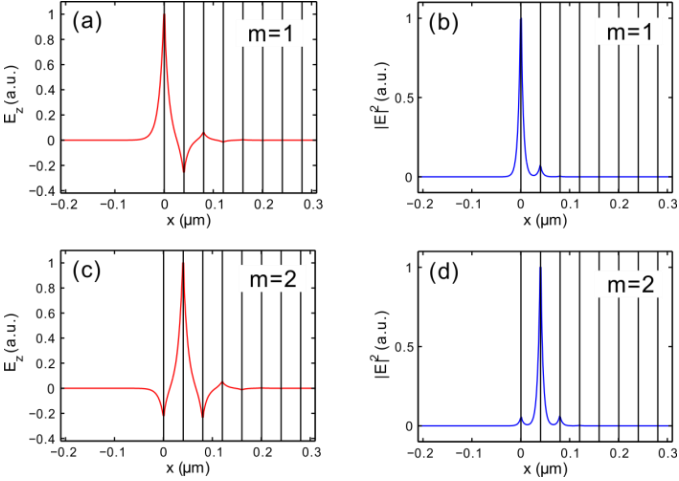

Fig. 1. (a) (b) The normalized tangential electric field ($E_z$) and the normalized intensity distribution ($|E|^2$) of the surface PLSs for the $m = 1$ mode. (c) (d) The normalized tangential electric field ($E_z$) and the normalized intensity distribution ($|E|^2$) of the surface PLSs for the $m = 2$ mode. The black lines stand for the graphene sheets which embedding in the dielectric represented by the blank.

Taking the graphene nonlinearity into consideration, Eq. (1) turns into a nonlinear eigen problem, which could be solved by using the self-consistent method [31]. We set a Gaussian distribution around the array boundary as initial value for the iterative calculation, where the peak intensity is set as 200 $V^2/\mu m^2$. The transverse field distribution of surface PLSs, namely, the nonlinear eigenmode profile, is obtained as shown in Fig. 1. Here, the loss of graphene is not taken into account at first. It has also been verified that the solutions do not change much even though the loss is included. We denote the nonlinear eigenmode as $m = 1$ or 2 mode when the peak of the intensity profile localizes at the first or second graphene sheet. For the mode of $m = 1$, the tangential electric field ($E_z$) and the intensity distribution $|E|^2 = |E_x|^2 + |E_z|^2$ are presented in Figs. 1(a) and 1(b). The transverse field distribution of surface PLS is asymmetric as shown in Fig. 1(b). Due to the boundary effect, the attenuation degree of the intensity amplitude in the uniform dielectric region is a little larger than that in the array. The power and the effective width of surface PLS are respectively given by [13,32]

$$P = \frac{1}{2}\int \text{Re}(E_xH_y^*)dx, \quad (3)$$

$$w = \sqrt{\int x^2|E|^2 dx / \int |E|^2 dx}. \quad (4)$$

For the mode $m = 1$, the surface PLSs have a width of 0.013 μm which is equal to $0.0013\lambda$, while the input power is about 19 W/m. As for the $m = 2$ mode, the tangential electric field and the intensity distribution are respectively shown in Figs. 1(c) and 1(d). Compared with the $m = 1$ mode, the intensity distribution of $m = 2$ mode tends to be symmetric with the intensity peak moving to deeper lattice region. In Fig. 1(d), the effective width of the surface PLS ($m = 2$) is about $0.0043\lambda$ which is almost three times larger than that of $m = 1$ mode. Meanwhile, the soliton power which equals 20.1 W/m is a little larger than that of $m = 1$ mode. For the same peak intensity, the surface PLS energy could be more concentrated as the distribution is closer to the array boundary. In general, the transverse size of the surface PLS could be squeezed into deep-subwavelength scale ($\sim 0.001\lambda$) with relatively low incident power.

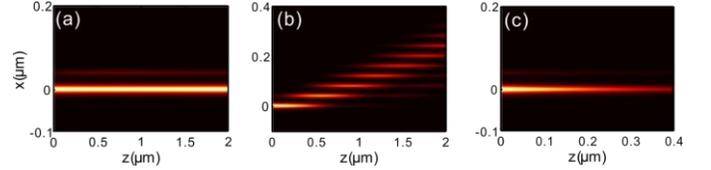

Fig. 2. (a) Propagation of the surface PLSs in the lossless semi-infinite graphene sheet arrays. (b) Diffraction of SPPs as the nonlinearity is not considered. (c) Actual propagation of the surface PLSs in the lossy semi-infinite graphene sheet arrays.

To illustrate the propagation of the surface plasmonic lattice solitons, we substitute the nonlinear eigenmode solution into the full Maxwell's equation and simulate the propagation of the mode by using the modified split-step Fourier beam propagation method [12]. In the simulation, we focus on the $m = 1$ mode. As shown in Fig. 2(a), the surface PLS keeps its initial distribution and propagates forward stably, where the loss from graphene is neglected. Assuming that the graphene nonlinear effect is not excited, the light beam will diffuse into the graphene sheet arrays, which is shown in Fig. 2(b). The surface solitons are formed due to the balance between graphene nonlinearity and the SPPs tunneling. Taking the loss from graphene into consideration, we also plot the real propagation of the surface PLS as shown in Fig. 2(c). The decay distance $L_{loss}$ is about 0.28 μm with the propagation constant $k_z = 106.89 + 1.76i$ μm$^{-1}$ ($L_{loss} = 1/2\text{Im}(k_z)$). The corresponding SPP wavelength is $\lambda_p = 0.059$ μm. Accordingly, the decay distance is almost five times of the SPP wavelength $L_{loss} = 4.7\lambda_p$.

### III. EXCITATION THRESHOLD OF SURFACE PLSs

Now we investigate the properties of surface PLSs including the threshold of exciting power and energy distribution. Firstly, we present the relation between the input power and the propagation constant of surface PLSs in Fig. 3(a), where the blue (red) line corresponds to the $m = 1$ ($m = 2$) mode. In Fig. 3(a), the threshold power of $m = 1$ mode is about 16.5 W/m.

Below the value, there is no solution for the surface PLSs which concentrate at the first graphene sheet. For the $m = 2$ mode, the threshold power is about 15 W/m, which is smaller than that of the $m = 1$ mode. Combined with the intensity distributions of $m = 1$ and 2 modes as shown in Figs. 1(b) and 1(d), we find that the threshold power would decrease as the surface PLS distribution gradually shifts to the internal graphene sheets.

Concerning on the surface PLSs of $m = 1$ mode, we proceed to analyze the influence of the light intensity on the energy distribution. The result is shown in Fig. 3(b), where $r_1$ and $r_2$ represent the proportions of the surface PLS energy confined in the first and second graphene sheet, respectively. In Fig. 3(b), it is clear that the soliton energy is more concentrated on the first graphene sheet as the peak intensity increases. Specially, when the intensity peak is larger than 200 $V^2/\mu m^2$, there is more than ninety percent of total energy is confined in the first graphene sheet. At the same time, it also indicates that the effective width is correspondingly less than $0.001\lambda$. When the light intensity becomes larger, the induced nonlinear effect will be enhanced. As a result, the surface PLSs could be further squeezed with more energy concentrated on the first graphene sheet.

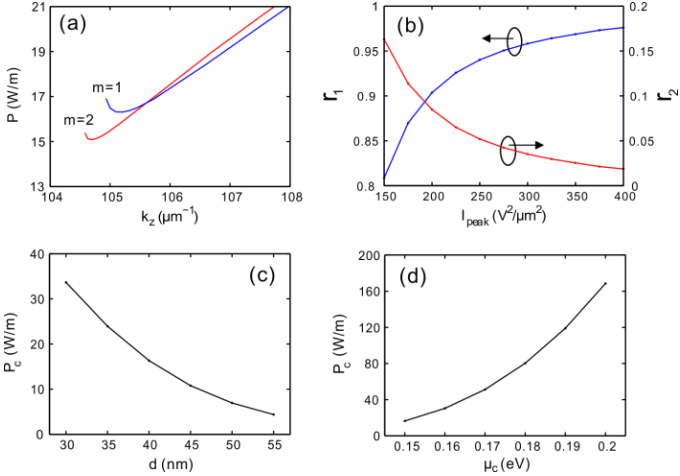

Fig. 3. (a) The relation between the input power and the propagation constant. (b) For $m = 1$ mode, the proportion of the energy confined around the first graphene sheet (blue line) and the second sheet (red line). (c) (d) The threshold power of the $m = 1$ mode versus the period of the structure and the chemical potential of graphene.

We also investigate the influence of the structure period and the chemical potential on the threshold power of the $m = 1$ mode. When the chemical potential is fixed at $\mu_c = 0.15$ eV, the threshold power decreases as the structure period increases, as shown in Fig. 3(c). In case the spacing between adjacent graphene sheets increases, the SPPs coupling becomes weak which needs relatively low nonlinear effect to balance. Consequently, the minimum power which is enough to give rise to the surface PLSs should be reduced. On the other hand, due to the great influence of the chemical potential on the material characteristics of the semi-infinite GSAs, we also reveal the influence of the threshold power on the chemical potential of graphene. The result is shown in Fig. 3(d), where the structure period is fixed at $d = 40$ nm. As the chemical potential is modulated in a small range of $\mu_c = 0.15 \sim 0.2$ eV, the threshold power will vary dramatically from 16.5 W/m to 168.5 W/m. The threshold power is very sensitive to the chemical potential. In general, the larger chemical potential would enhance the SPPs coupling which requires stronger nonlinear effect to balance [18,21]. Thus, the corresponding threshold power becomes larger.

## IV. APPLICATION IN OPTICAL SWITCHES

In this part, we shall discuss about the application in optical switches by manipulating the SPPs propagation in the semi-infinite GSAs. At first, we simulate the stable propagation of the surface PLS which is extremely concentrated at the outmost graphene sheet with the peak intensity $I_{peak} = 350$ $V^2/\mu m^2$. The same illustrations of the surface PLS propagation are presented in Figs. 4 (a) and 5 (a), where the chemical potential and the structure period are set as $\mu_c = 0.15$ eV and $d = 40$ nm, respectively. Then we demonstrate how the light beam diffracts from the boundary into the inner graphene sheet arrays as the balance between nonlinearity and SPPs tunneling is destroyed by changing the chemical potential or the input power. The results are shown in Figs. 4 and 5.

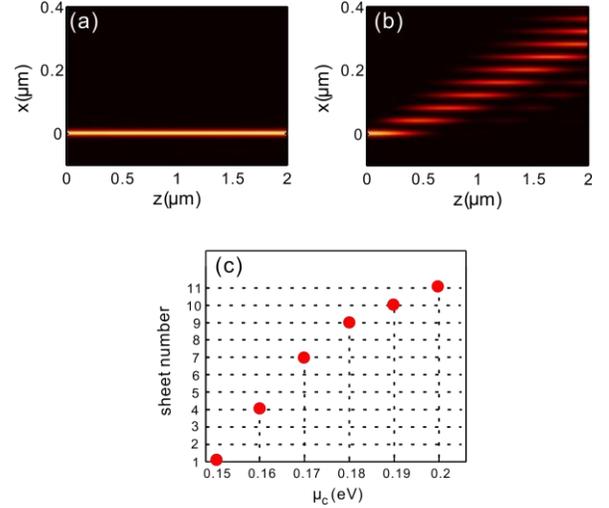

Fig. 4. (a) The surface PLS propagates along the first graphene sheet with the peak intensity $I_{peak} = 350$ $V^2/\mu m^2$ as the chemical potential is 0.15 eV. (b) When the chemical potential is modulated from $\mu_c = 0.15$ eV to $\mu_c = 0.18$ eV, the light beam diffuses into the internal graphene sheets. (c) As the chemical potential increases, the light output position becomes farther away from the array boundary at the distance $z = 2$ μm.

It is known that the chemical potential of graphene can be tuned flexibly by external electric field, magnetic field or gate voltage [21,22]. As the chemical potential is increased from $\mu_c = 0.15$ eV to $\mu_c = 0.18$ eV, the light beam is diffracted into the array as shown in Fig. 4(b). In the previous part, we have discussed that the increasing chemical potential leads to stronger SPPs coupling. It also indicates that the balance between the graphene nonlinearity and SPP tunneling is broken and the linear diffraction dominates. In Fig. 4(b), the peak of the intensity profile localizes at the 9*th* graphene sheet ($x = 0.32$ μm) when the light beam arrives at the output end $z = 2$ μm. To illustrate the influence of the chemical potential on the light output position in more detail, we increase the chemical potential from $\mu_c = 0.15$ eV to $\mu_c = 0.2$ eV successively. The corresponding output positions of the intensity peak are illustrated in Fig. 4(c). As the chemical potential increases, the diffraction becomes in domination and the output location stays farther away from the array boundary. By tuning the chemical potential even in the small range of $\mu_c$ = 0.15 ~ 0.2 eV, ones can flexibly manipulate the output location of the light.

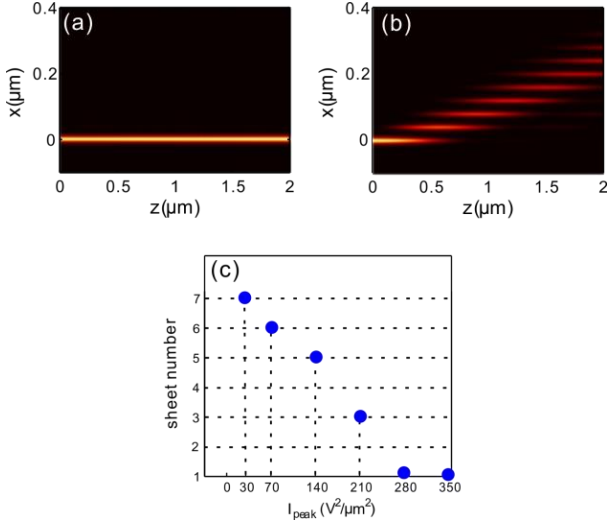

Fig. 5. (a) The surface PLS propagates along the first graphene sheet with the peak intensity $I_{peak}$ = 350 V$^2$/μm$^2$ as the chemical potential is 0.15 eV. (b) The light beam diffuses into the inner sheets of the array as the peak intensity is reduced to $I_{peak}$ = 30 V$^2$/μm$^2$. (c) As the peak intensity decreases, the output localization of the intensity becomes farther away from the array boundary at the output end $z$ = 2 μm.

The influence of the input power on the output position of light is also studied. Based on the stable propagation of the surface PLS as shown in Fig. 5(a), we reduce the peak intensity from $I_{peak}$ = 350 V$^2$/μm$^2$ to $I_{peak}$ = 30 V$^2$/μm$^2$, meanwhile the power is also reduced. As a result, the light beam would diffuse into the inner sheets as shown in Fig. 5(b). At the output end $z$ = 2 μm, the intensity peak localizes at the 7$th$ graphene sheet. To get more information about how the input power dose influence on the output position, we reduce the peak intensity in turn from $I_{peak}$ = 350 V$^2$/μm$^2$ to $I_{peak}$ = 30 V$^2$/μm$^2$. The result is shown in Fig. 5(c). As the peak intensity decreases, the distance between the output position and the array boundary becomes larger. This is because that the light diffraction dominates as the nonlinear effect is weakened by reducing the input power. In Fig. 5(c), it should be noted that although the intensity peak localizes at the first sheet as the peak intensity is reduced to $I_{peak}$ = 280 V$^2$/μm$^2$, quite a few energy has diffused into the second and third sheets at the output facet. By tuning the input power, we could also control the output position of the light.

## V. CONCLUSION

In conclusion, we have investigated the surface plasmonic lattice solitons in semi-infinite graphene sheet arrays. With the prerequisite that the light power is above a threshold value, surface PLSs will be formed as the SPPs tunneling and the graphene nonlinearity reach a balance. Due to the strong confinement of SPPs on graphene, the width of the surface PLSs could reach deep-subwavelength scale (~ 0.001$λ$) with relatively low incident power. The threshold power of surface PLSs can be reduced by increasing the structure period or decreasing the graphene chemical potential. By tuning the chemical potential or the input power, we can also flexibly manipulate the output position of light beam. The study would provide potential application in optical switches on deep-subwavelength scale.

ACKNOWLEDGMENT

This work is supported by the 973 Program (No. 2014CB921301), the National Natural Science Foundation of China (Nos. 11304108, 11674117), Natural Science Foundation of Hubei Province (2015CFA040), and the Specialized Research Fund for the Doctoral Program of Higher Education of China (No. 20130142120091).

(Corresponding author: B. Wang
e-mail: wangbing@hust.edu.cn).
(Corresponding author: P. Lu
e-mail: lupeixiang@hust.edu.cn).

Z. Wang, B. Wang, H. Long, and K. Wang are with School of Physics and Wuhan National Laboratory for Optoelectronics, Huazhong University of Science and Technology, Wuhan 430074, China. (E-mail: wangbing@hust.edu.cn).

P. Lu is with School of Physics and Wuhan National Laboratory for Optoelectronics, Huazhong University of Science and Technology, Wuhan 430074, China, and also with the Laboratory of Optical Information Technology, Wuhan Institute of Technology, Wuhan 430205, China (E-mail: lupeixiang@hust.edu.cn).